\documentclass[11pt, a4paper, twocolumn]{article}
\pdfoutput=1
\usepackage[utf8]{inputenc}
\usepackage[T1]{fontenc}
\usepackage{lmodern, textcomp}
\usepackage[american]{babel}
\usepackage{csquotes}

\usepackage{eurosym}
\usepackage[nottoc]{tocbibind}
\usepackage{authblk}
\usepackage{fancyhdr}
\usepackage{setspace}
\usepackage{titlesec}
\usepackage{changepage}
\usepackage{lastpage}
\usepackage{xspace}

\usepackage{graphicx}
\usepackage{subcaption}
\usepackage{float}
\usepackage{makecell}
\usepackage{booktabs}
\usepackage{multirow}
\usepackage{marginnote}
\usepackage{enumitem}
\usepackage[multiple]{footmisc}

\usepackage[usenames, dvipsnames, svgnames, table]{xcolor}

\usepackage[backend=bibtex8, citestyle=numeric-comp, maxbibnames=99,
			sorting=none, sortcase=false, sortcites=true, giveninits=true]{biblatex}

\usepackage{amsmath, amsfonts, amssymb}
\usepackage{mathtools}
\usepackage{cancel}
\usepackage[separate-uncertainty=true]{siunitx}
\usepackage{listings}

\usepackage[pdftex, breaklinks=true, linktocpage,
	colorlinks=true, urlcolor=blue, linkcolor=blue, citecolor=red
]{hyperref}

\usepackage{abstract}
\usepackage{caption}

\newcommand{\email}[1]{\href{mailto:#1}{\nolinkurl{#1}}}
\newcommand{\emailfoot}[1]{\thanks{\email{#1}}}

\newcounter{draftcommentcnt}
\NewDocumentCommand{\draftcomment}{s O{red} m}{%
	\def\margnote{\IfBooleanTF{#1}{\marginnote}{\marginpar}}%
	\stepcounter{draftcommentcnt}%
	\textcolor{#2}{#3}%
	\margnote{\textcolor{#2}{$\Leftarrow$ \arabic{draftcommentcnt}}}%
}

\numberwithin{equation}{section}

\usepackage[sort&compress, english]{cleveref}

\usepackage[heightrounded, margin=0.75in,footskip=0.25in]{geometry}
\newcommand{\preprint}[0]{MIT-CTP/5319 \\ UUITP-36/21}
\fancypagestyle{preprint}{
	\fancyhf{}

	\fancyhead[R]{\small\textsf{\preprint}}
}

\input{arxiv.def}

\hypersetup{pdftitle={Deep multi-task mining Calabi-Yau four-folds}}

\title{\bf Deep multi-task mining Calabi-Yau four-folds}

\author[1,2,3]{Harold Erbin\emailfoot{erbin@mit.edu}}
\author[3,4]{Riccardo Finotello\emailfoot{riccardo.finotello@cea.fr}}
\author[5]{Robin Schneider\emailfoot{robin.schneider@physics.uu.se}}
\author[3]{Mohamed Tamaazousti\emailfoot{mohamed.tamaazousti@cea.fr}}

\affil[1]{%
	Center for Theoretical Physics, Massachusetts Institute of Technology
	\protect\\
	Cambridge, MA 02139, \textsc{Usa}
}

\affil[2]{%
	\textsc{Nsf Ai} Institute for Artificial Intelligence and Fundamental Interactions
}

\affil[3]{%
	Université Paris Saclay, \textsc{Cea}, \textsc{List}
	\protect\\
	Palaiseau, F-91120, France
}

\affil[4]{%
	Université Paris Saclay, \textsc{Cea}, Service d'Études Analytiques et de Réactivité des Surfaces (SEARS)
	\protect\\
	Gif-sur-Yvette, F-91191, France
}

\affil[5]{%
    Department of Physics and Astronomy, Uppsala University
    \protect\\
    SE-751 20 Uppsala, Sweden
}

\begin{document}

\twocolumn[
    \maketitle
    \begin{onecolabstract}
    We continue earlier efforts in computing the dimensions of tangent space cohomologies of Calabi-Yau manifolds using deep learning.
    In this paper, we consider the dataset of all Calabi-Yau four-folds constructed as complete intersections in products of projective spaces.
    Employing neural networks inspired by state-of-the-art computer vision architectures, we improve earlier benchmarks and demonstrate that all four non-trivial Hodge numbers can be learned at the same time using a multi-task architecture.
    With \SI{30}{\percent} (\SI{80}{\percent}) training ratio, we reach an accuracy of \SI{100}{\percent} for $h^{(1,1)}$ and \SI{97}{\percent} for $h^{(2,1)}$ (\SI{100}{\percent} for both), \SI{81}{\percent} (\SI{96}{\percent}) for $h^{(3,1)}$, and \SI{49}{\percent} (\SI{83}{\percent}) for $h^{(2,2)}$.
    Assuming that the Euler number is known, as it is easy to compute, and taking into account the linear constraint arising from index computations, we get \SI{100}{\percent} total accuracy.
    \end{onecolabstract}
	\vspace{0.5cm}

    \thispagestyle{preprint}
]

\saythanks

\hrule
\pdfbookmark[1]{\contentsname}{toc}
\tableofcontents
\bigskip
\hrule
\bigskip

\section{Introduction}
\label{sec:intro}

There is a growing body of research that applies modern techniques from data science to problems in string theory~\cite{Ruehle:2020jrk}. The reasons for that are two-fold. On the one hand, standard computations in string theory are hard, in particular they can be NP-hard or even undecidable~\cite{Ruehle:2020jrk,Denef:2007:ComputationalComplexityLandscape,Halverson:2018cio}. Due to double exponential scaling laws in terms of computational resources with respect to the input parameters, string theory calculations often fail to finish in a reasonable amount of time even on modern machines.
On the other hand, there are too many configurations to consider. The largest estimates put a bound of $\mathcal{O}(\num{e272000})$ when considering F-theory compactified on a Calabi-Yau four-fold~\cite{Taylor:2015xtz}. Parsing that many configurations is impossible, thus computational smart ways are needed to select potentially interesting vacuum configurations~\cite{Halverson:2019tkf,Larfors:2020ugo}.

An important key component in realistic string theory compactifications are Calabi-Yau manifolds. These manifolds have been studied extensively in the past, and thus they comprise some of the best datasets within the string theory community~\cite{He:2020:CalabiYauSpacesString}:
\begin{enumerate}
    \item The first widely used dataset are the \num{7 890} complete intersection Calabi-Yau, in short CICY, manifolds in three complex dimensions by Candelas et al.~\cite{Candelas:1987kf,Green:1987cr,Anderson:2017:FibrationsCICYThreefolds}.
    
    \item The largest dataset, the Kreuzer-Skarke list, contains 473 million reflexive polytopes in four dimensions. These encode a toric ambient space, from which one obtains Calabi-Yau three-folds by considering the hypersurface defined by the canonical bundle~\cite{Kreuzer:2000xy}.
    
    \item CICY four-folds have also been classified and amount to \num{921 497} distinct configuration matrices~\cite{Gray:2013mja,Gray:2014fla}.
\end{enumerate}

The incredible progress in data science, in particular image recognition, over the past decade can in part be attributed to large and clean datasets~\cite{ILSVRC15}. They allowed researchers to benchmark their algorithms and let the best ones compete against each other, which in turn resulted in rapid development and ever improving neural network architectures~\cite{10.1145/3065386,Szegedy:2015:GoingDeeperConvolutions,Szegedy:2016:RethinkingInceptionArchitecture,Szegedy:2017:Inceptionv4InceptionResNetImpact,he2016deep}.
We will proceed in a similar vein in this paper. The number of independent Kähler moduli of CICY three-folds has been successfully analyzed using neural networks in the past. These benchmarks were initiated by He, who proposed to treat their configuration matrices as a simple two-dimensional image~\cite{He:2017aed}.

In previous works~\cite{Erbin:2020srm,Erbin:2020tks}, two of the authors have shown that learning $h^{(1,1)}$ is possible to great accuracy, but the limited training data is not sufficient to generalize the learning to the number of complex structure moduli $h^{(2,1)}$.
Computing Hodge numbers of Calabi-Yau manifolds is of great importance, since cohomology computations are an integral part of string theory compactifications. They for example determine the number of massless fermion generations in string theory compactifications. Thus, the goal is to identify performant algorithms in these well-studied datasets of tangent bundle cohomologies, which can then generalize to more complicated vector bundles.

In the rest of this paper, we will present two different approaches for learning Hodge numbers of CICY four-folds. First, we will treat the problem as a standard image classification task where the Hodge numbers are the image labels. For this purpose, we employ an Inception module based architecture~\cite{Szegedy:2015:GoingDeeperConvolutions,Szegedy:2016:RethinkingInceptionArchitecture,Szegedy:2017:Inceptionv4InceptionResNetImpact,Erbin:2020srm,Erbin:2020tks} and show that a single set of hyperparameters generalizes well to all four Hodge numbers, yielding a mean accuracy over all Hodge numbers of \SI{85}{\percent}. This suggests that we could scale the approach to a multi-task learning problem.

Subsequently, we show that all Hodge numbers can be learned simultaneously by utilizing a branched network with hard parameter sharing~\cite{Caruana93multitasklearning:, standley2020tasks} between the task specific sub-structures, which ultimately are responsible for learning the distributions of the Hodge numbers.
The multi-task approach has several advantages, with respect to single-task architectures. From a technical side, multi-task learning has been shown to improve the overall performance of the models~\cite{Caruana93multitasklearning:}. From a physics and algebraic geometry perspective, a single model hints towards the definition of a unified framework from which it may be possible to extract meaningful theoretical information, such as closed form formulas.
The model we developed is capable of learning at the same time, and without rescaling, the four dimensions of the tangent space cohomologies of CICYs, accounting for the heavy class imbalance present in the dataset.
This multi-task Ansatz leads to perfect performance on two of the four Hodge numbers and accuracy of \SI{96}{\percent} and \SI{83}{\percent} for $h^{(3,1)}$ and $h^{(2,2)}$ respectively, with a training ratio of \SI{80}{\percent}.

The outline of this paper is as follows. In \Cref{sec:related}, we discuss related works of learning cohomologies and earlier results on Calabi-Yau three-folds. \Cref{sec:dataset} explores the dataset of CICY four-folds and presents the results of our classification experiments. This is followed by our main results in \Cref{sec:results} in which we introduce our deep learning model \emph{CICYMiner}, a multi-task regression model based on chained Inception modules that predicts all four Hodge numbers at once. We conclude in \Cref{sec:outlook} with some outlooks.
Python codes for this paper can be found at:
\begin{itemize}
    \item \url{https://github.com/robin-schneider/cicy-fourfolds}
    \item \url{https://github.com/thesfinox/ml-cicy-4folds}
\end{itemize}
The list of packages used throughout the development comprises \texttt{pandas}~\cite{reback2020pandas, mckinney-proc-scipy-2010} and \texttt{numpy}~\cite{harris2020array} for data operations, \texttt{matplotlib}~\cite{Hunter:2007} and \texttt{seaborn}~\cite{Waskom2021} for visualisation, and \texttt{tensorflow}~\cite{tensorflow2015-whitepaper} for the deep learning algorithms.

\begin{figure*}[t]
    \centering
    \begin{minipage}{0.47\textwidth}
    \includegraphics[width=\textwidth]{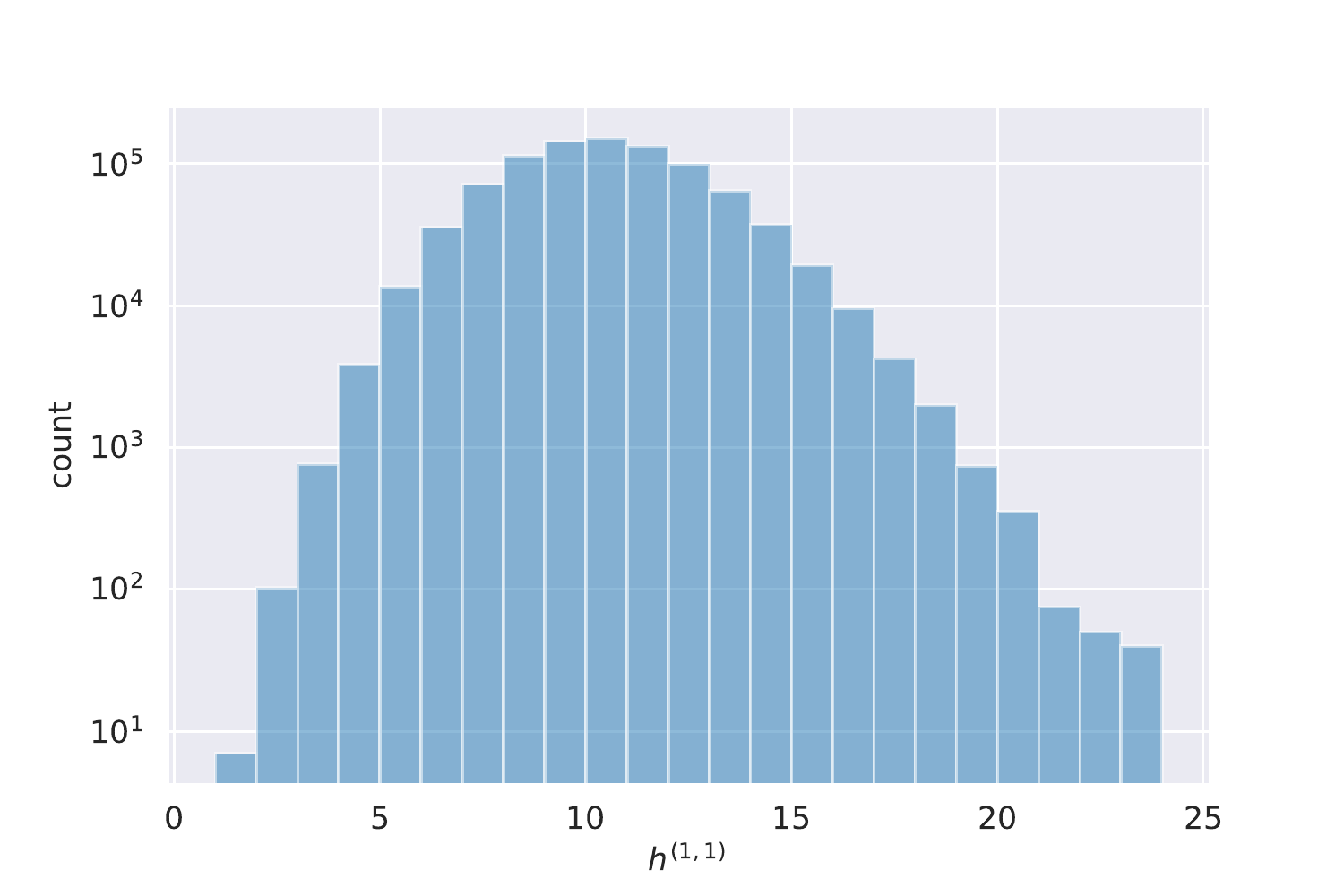}
    \includegraphics[width=\textwidth]{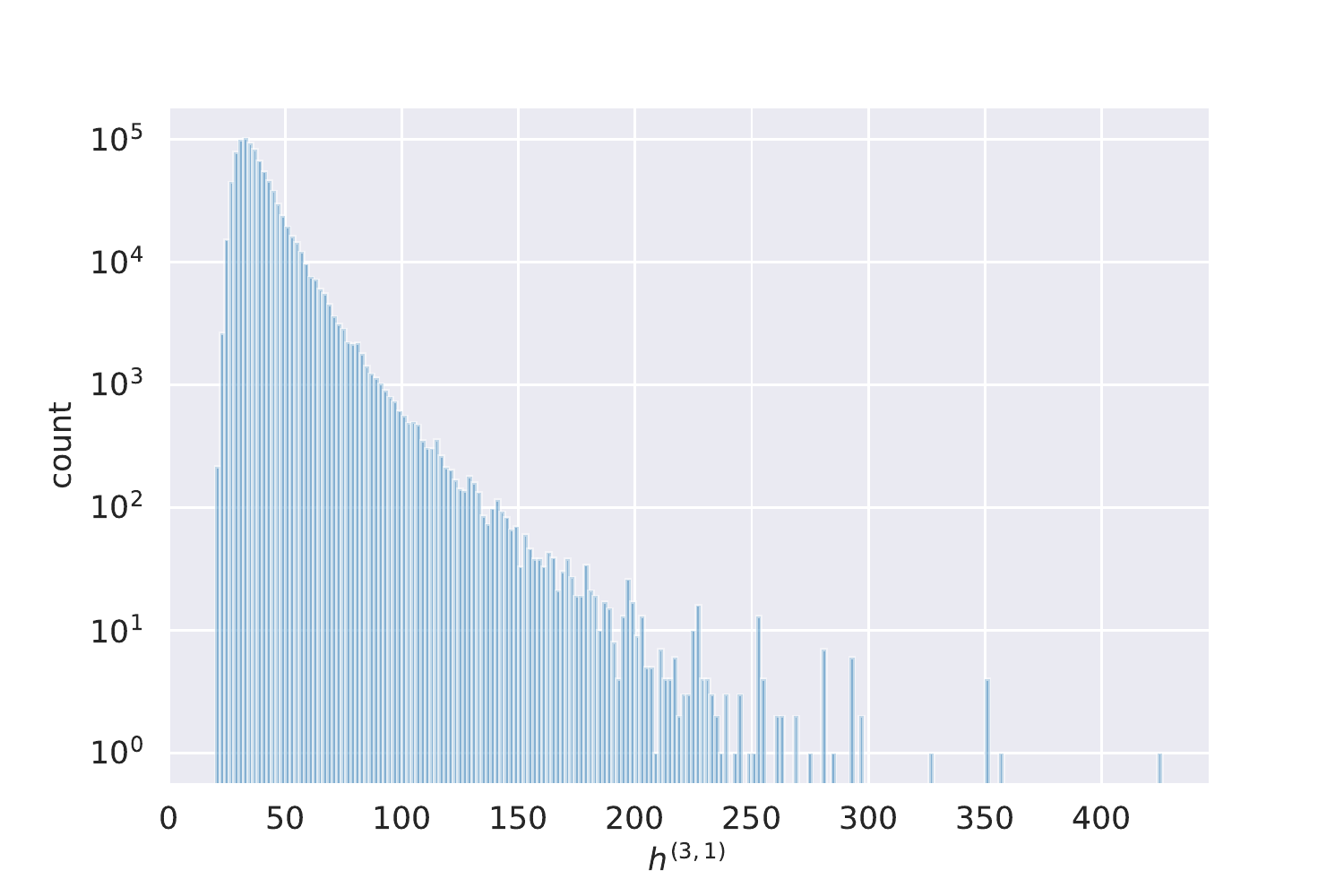}
    \end{minipage}
    \begin{minipage}{0.47\textwidth}
    \includegraphics[width=\textwidth]{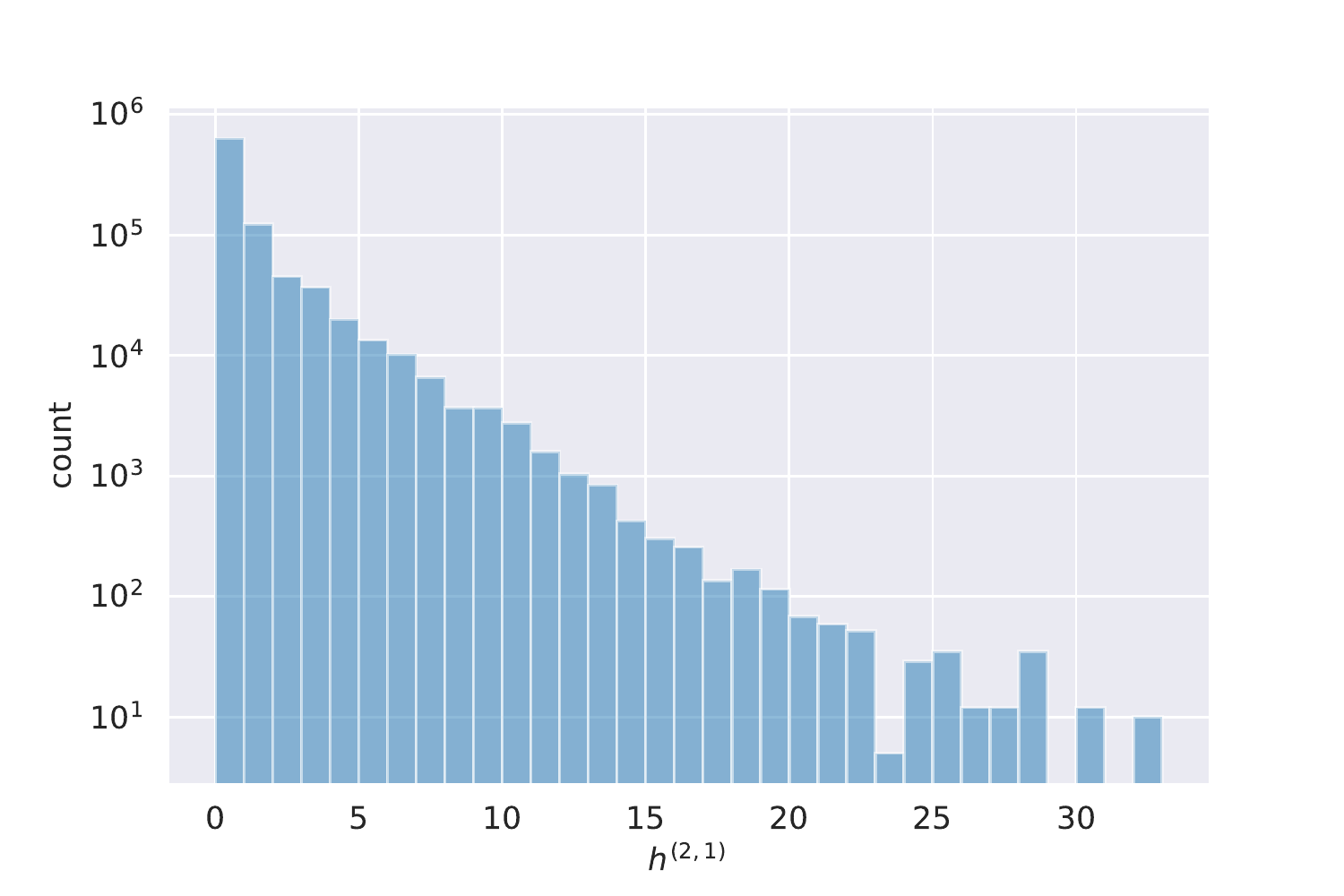}
    \includegraphics[width=\textwidth]{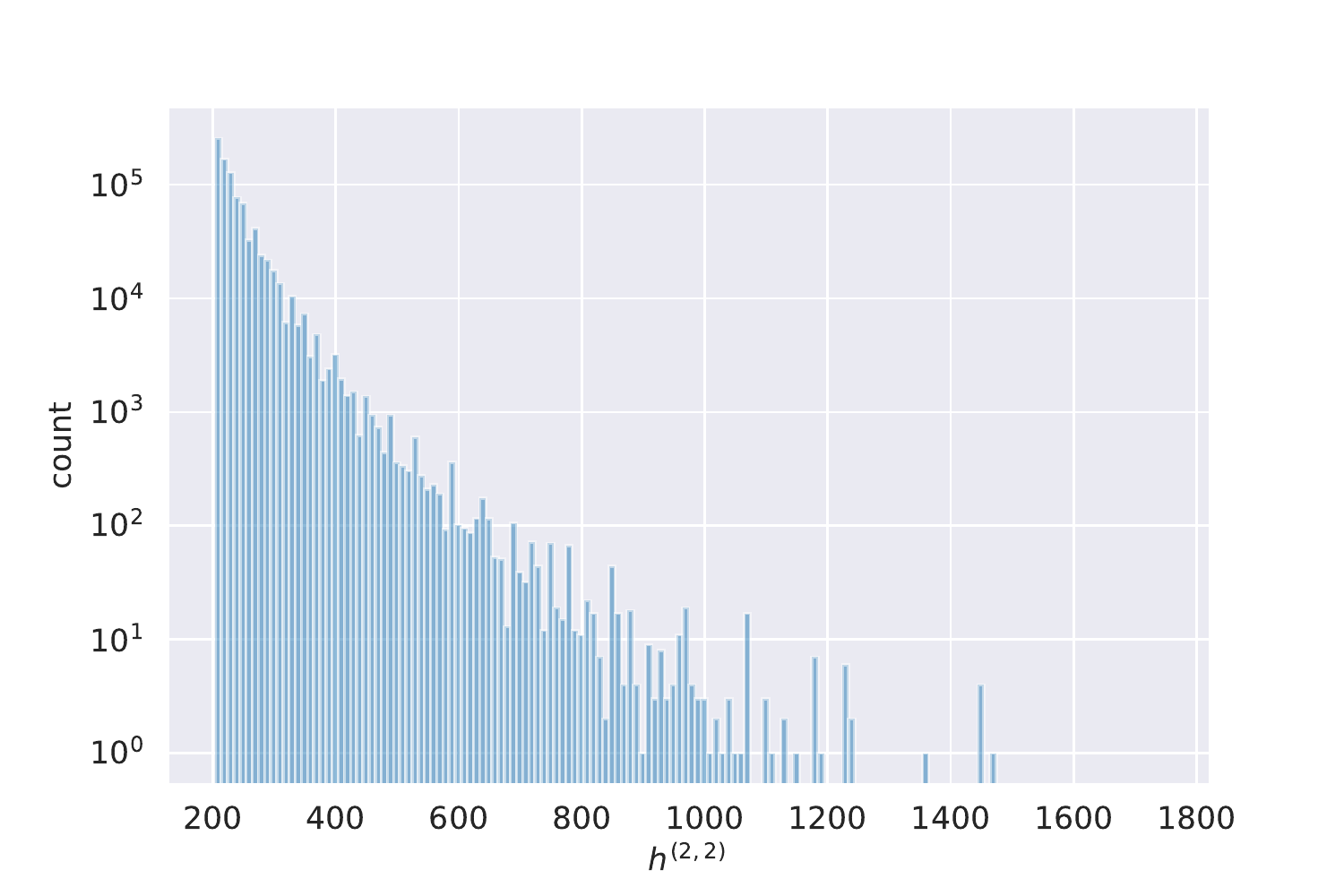}
    \end{minipage}
    \caption{\it The plots show the histograms with logarithmic $y$-axis of the four non-trivial Hodge numbers. In the first row we have on the left the distribution of $h^{(1,1)}$, to the right of $h^{(2,1)}$. In the bottom row $h^{(3,1)}$ is presented to the left and $h^{(2,2)}$ to the right. }
    \label{fig:Hodgehist}
\end{figure*}

\section{Related works}
\label{sec:related}


The first paper utilizing machine learning algorithm to predict various different cohomology dimensions was written by He~\cite{He:2017aed}.
The author tackled the problem of predicting Hodge numbers of CICY three- and four-folds, but also line bundles over these manifolds~\cite{He:2017aed}. These studies have later been extended to systematically investigate CICY three-folds with linear regression, support vector machines, and dense neural networks achieving accuracies ranging from \SI{37}{\percent} to \SI{85}{\percent}~\cite{Bull:2018uow,Bull:2019cij} when using \SI{70}{\percent} training data. The benchmarks have subsequently been improved by using an Inception-based architecture to accurately predict \SI{97}{\percent} of the test data using only \SI{30}{\percent} training data, essentially solving the problem of predicting $h^{(1,1)}$~\cite{Erbin:2020srm,Erbin:2020tks}. This work was supplemented by more methodological studies in which the dataset was augmented with various other (topological) quantities.
Other works on CICY three-folds include ~\cite{Krippendorf:2020:DetectingSymmetriesNeural,He:2019:DistinguishingEllipticFibrations}.

An initial exploration of CICY four-folds has been started by He and Lukas~\cite{He:2017aed,He:2020lbz}. The authors used a simple dense neural network and were able to predict $h^{(1,1)}$ with an accuracy of \SI{96}{\percent}. This promising early result showed that the increased size of the dataset improves the performance significantly. However, in line with previous studies of $h^{(2,1)}$ on CICY three-folds, the authors were unable to accurately predict the value of the other Hodge numbers, reaching an accuracy of only \SI{27}{\percent} for $h^{(3,1)}$. They were successful in improving this accuracy for a subset of the dataset by considering all configuration matrices of shape $(4,4)$ and using feature enhancement. This feat was achieved by supplementing the training samples with all up to degree four monomials of the defining polynomials and pushed the accuracy to \SI{95}{\percent}.

The Kreuzer-Skarke list has also been the target of deep learning algorithms. In order to identify equivalent Calabi-Yau manifolds coming from different triangulations, Demirtas et al. trained residual neural networks to learn the triple intersection numbers~\cite{Demirtas:2020dbm}. They reached an almost perfect performance, which allowed them to cut down the computation time from seconds to microseconds. This in turn made it possible to derive an upper bound on the number of distinct Calabi-Yau manifolds arising from the polytope with the most triangulations, setting it to $\num{E428}$.

There are several ongoing projects in learning Hodge numbers of line bundle cohomologies. These can be separated into two different approaches. First, learning the cohomology dimensions directly, for example on del Pezzo surfaces~\cite{Bies:2020gvf} and on CICY three-folds~\cite{He:2017aed,Ruehle:2017mzq,Larfors:2019sie,Larfors:2020ugo}. Second, neural networks have been used to classify cones in the cohomology-dimension landscape~\cite{Klaewer:2018sfl,Brodie:2019dfx,Brodie:2020:IndexFormulaeLine}. The Hodge numbers belonging to these cones can all be described by the same analytic equations~\cite{Constantin:2018hvl}.

\begin{figure*}[t]
    \centering
    \includegraphics[width=0.9\textwidth]{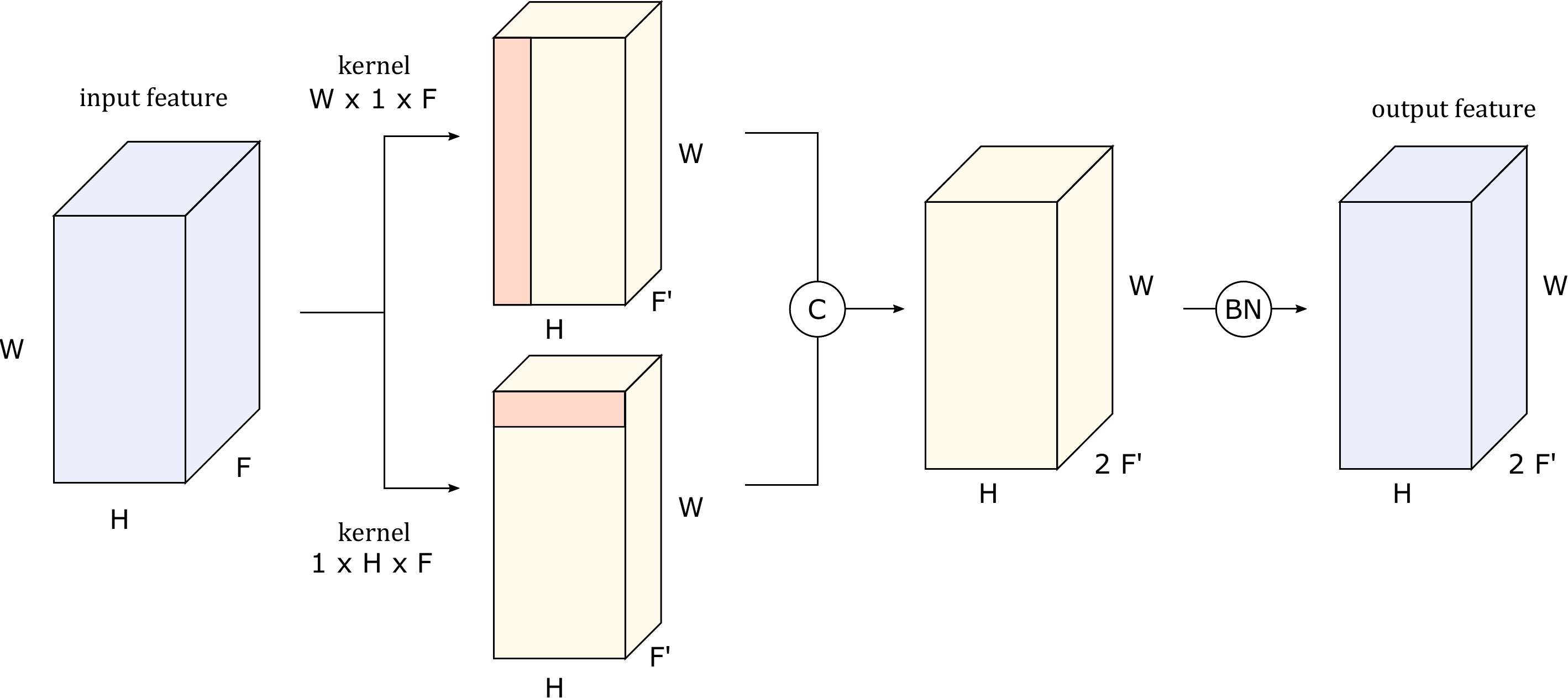}
    \caption{\it An Inception module can be decomposed into the different convolutional kernels scanning over the width (W) and height (H) with filters (F). They are subsequently concatenated (C) and followed by a batch normalization (BN) layer. The Inception module is the main building block of both the CICYMiner and the classification architectures.}
    \label{fig:Inception}
\end{figure*}

\begin{figure*}[t]
    \centering
    \includegraphics[width=0.95\textwidth]{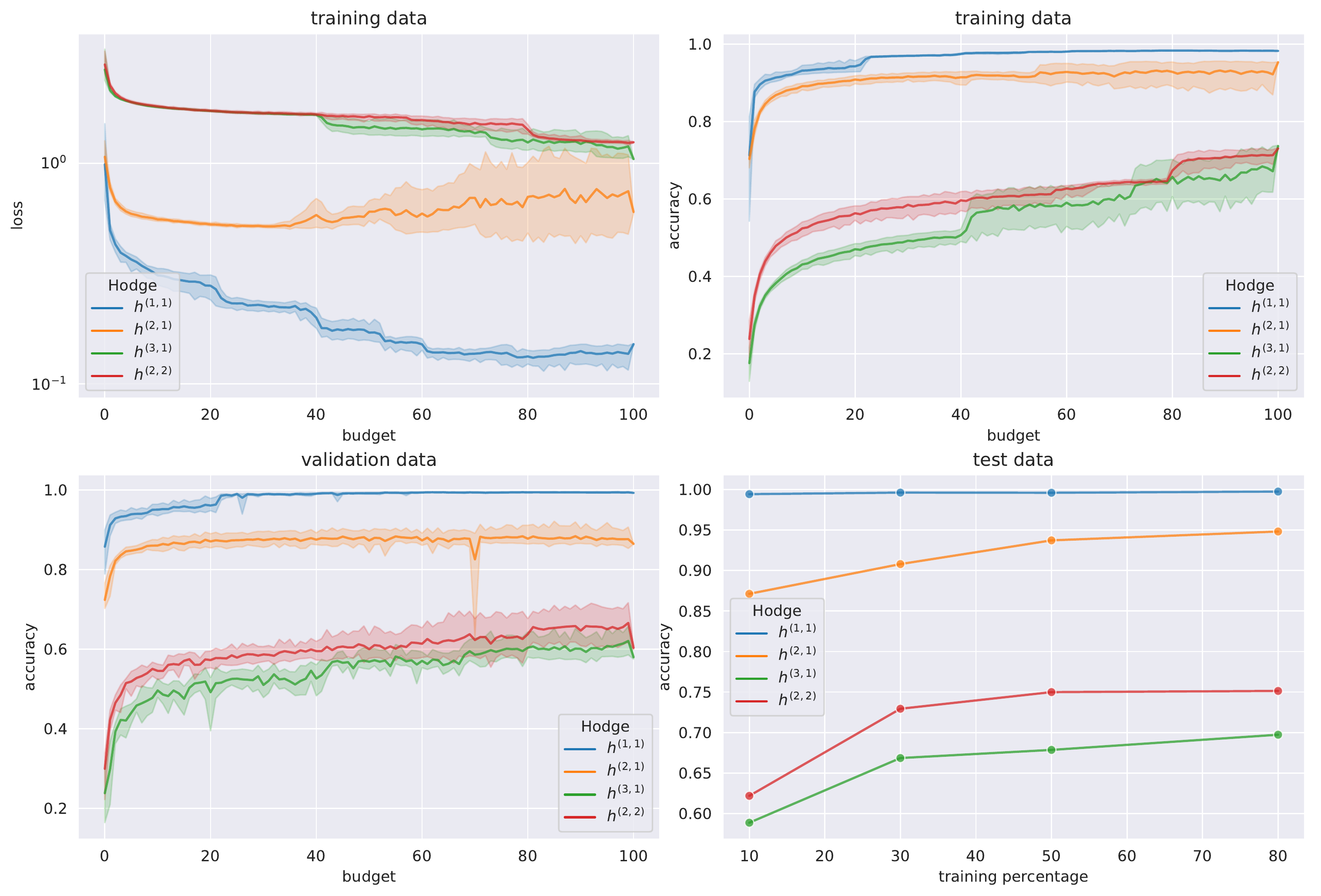}
    \caption{\it The first row shows training loss and accuracy plotted against the computation budget. The error bars represent the upper and lower bounds for the four different training ratios. In the second row, we plot on the left the validation accuracy and on the right the test accuracy of the best performing models for the different training ratios.}
    \label{fig:class}
\end{figure*}

\begin{figure*}[t]
    \centering
    \includegraphics[width=0.9\textwidth]{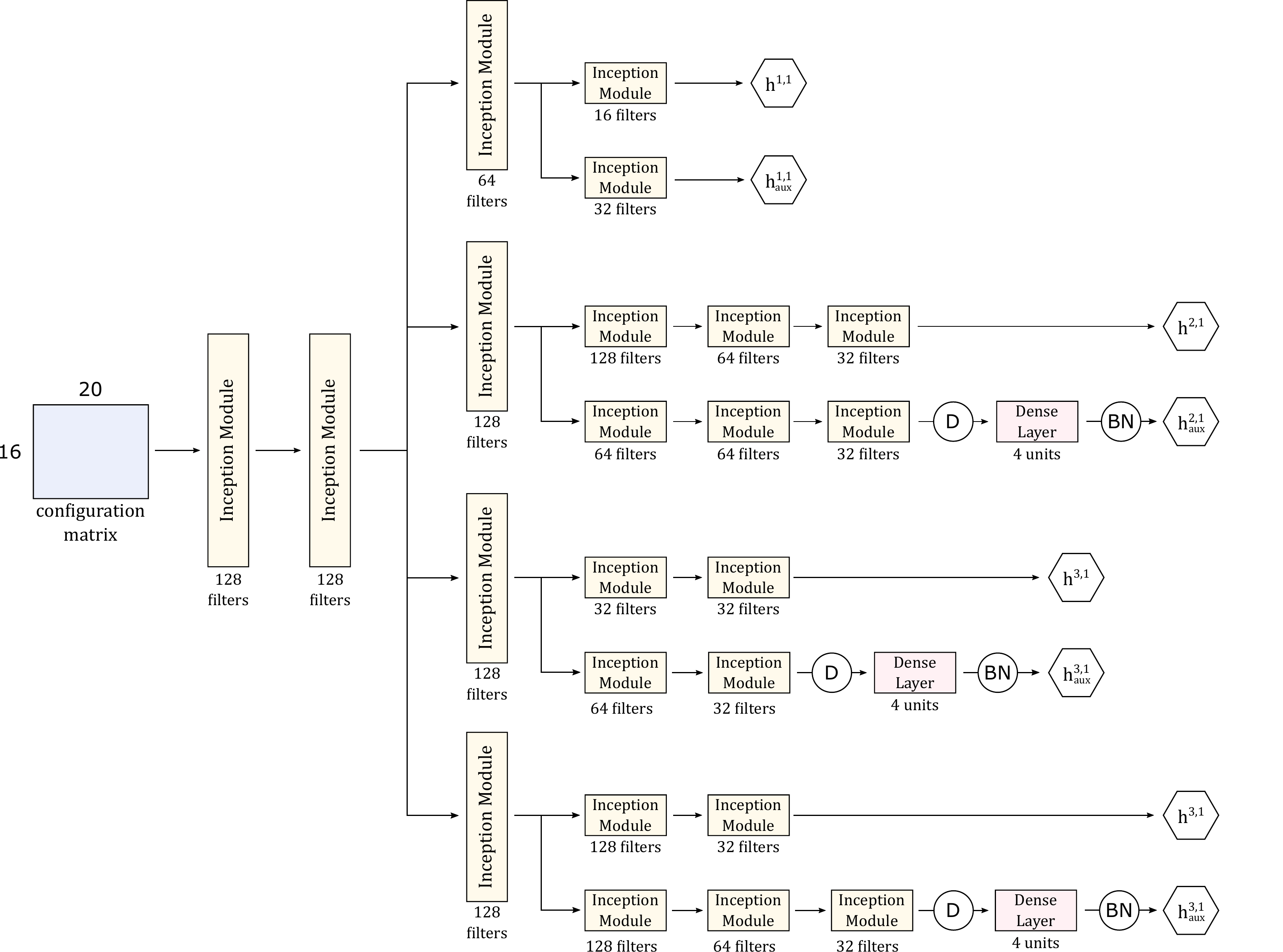}
    \caption{\it The basic building block of CICYMiner are Inception modules. The architecture is built to enable the hard parameter sharing in the bottom layer, in order to construct a common representation of the input. The task specific sub-structures then replicate the behaviour through an auxiliary branch, which further uses dense layers, batch normalization (BN) and dropout (D) to control overfitting. The final model predicts all Hodge numbers at once. The composition of the Inception modules are shown in \Cref{fig:Inception}.}
    \label{fig:deepminer}
\end{figure*}

\section{Exploring the dataset}
\label{sec:dataset}

In this section, we will introduce and explore complete intersection Calabi-Yau four-folds. We then proceed to learn the four non-trivial Hodge numbers independently using neural networks with an Inception inspired architecture~\cite{Szegedy:2015:GoingDeeperConvolutions,Szegedy:2016:RethinkingInceptionArchitecture,Szegedy:2017:Inceptionv4InceptionResNetImpact}.

\subsection{CICY four-folds}

A complete intersection Calabi-Yau manifold is fully defined by its configuration matrix. This matrix encodes the polynomial degrees and ambient space factors in the following way:
\begin{align}
\mathcal{M} =  \left[
\begin{array}{c||ccc}
n_0 & p^0_1 & \cdots & p^0_{K} \\
\vdots & \vdots & \ddots & \vdots \\
n_r & p^{r}_1 & \cdots & p^{r}_K  \\
\end{array}
\right]_{\chi}.
\end{align}
Each $p^i_j \in \mathbb{N}$ is the degree of the $j$-th polynomial in the homogeneous coordinates of the $i$-th complex projective space with dimension $n_i$. The Calabi-Yau condition is translated in the configuration matrix by requiring that
\begin{align}
n_i + 1 = \sum_{j=1}^K p^i_j.
\end{align}
The Euler number $\chi$ is given in the subscript and can be directly computed by integrating the fourth Chern class or from the four non-trivial Hodge numbers as
\begin{align}
\label{eq:euler}
    \chi = 4 + 2 h^{(1,1)} - 4h^{(2,1)} + 2 h^{(3,1)} + h^{(2,2)}.
\end{align}
A second linear relationship between the Hodge numbers can be derived by combining the indices $\chi_q = \chi(\mathcal{M}, \wedge^q T\mathcal{M}^*)$~\cite{Gray:2014fla} leading to
\begin{align}
\label{eq:linear}
    44 = - 4 h^{(1,1)} + 2 h^{(2,1)} - 4 h^{(3,1)} + h^{(2,2)}.
\end{align}
The configuration matrices have been generated from an initial set of matrices and subsequently applying the splitting procedure~\cite{Candelas:1987kf,Gray:2013mja}, finding new manifolds and discarding equivalent descriptions. In this way, a total of \num{921497} topological distinct types of CICY manifolds were found, with \num{905684} of them not being direct products of lower dimensional manifolds.

The Hodge number distributions are presented in \Cref{fig:Hodgehist}. The mean, maximum and minimum values are
\begin{align}
    & \langle h^{(1,1)} \rangle = 10.1^{24}_{1}, \quad \langle h^{(2,1)} \rangle = 0.817^{33}_{0}, \nonumber \\
    &\langle h^{(3,1)} \rangle = 39.6^{426}_{20}, \quad \langle h^{(2,2)} \rangle = 241^{1752}_{204}.
\end{align}
Notice that the distributions of the Hodge numbers are, in general, imbalanced: for instance, $h^{(2,1)}$ vanishes for \SI{70}{\percent} of the configuration matrices in the dataset.
We find that \SI{54.5}{\percent} are favourable (i.e.\ $h^{(1,1)}$ is equal to the number of projective spaces), less than the \SI{61.9}{\percent} for CICY three-folds.\footnote{There exists another dataset of CICY three-folds in which \SI{99.1}{\percent} are favourable~\cite{Anderson:2017:FibrationsCICYThreefolds}, but no such feature enhanced data is available for the four-folds. However, the results from~\cite{Erbin:2020srm,Erbin:2020tks} show that using favourable matrices helps mostly in computing $h^{(1,1)}$.}
Hence, for slightly more than half of the cases we have $h^{(1,1)} = r$, the number of projective ambient space factors. This number is important as it should be the baseline to compare any algorithm against.

\begin{figure*}[t]
    \centering
    \begin{minipage}{\textwidth}
    \includegraphics[width=0.47\textwidth]{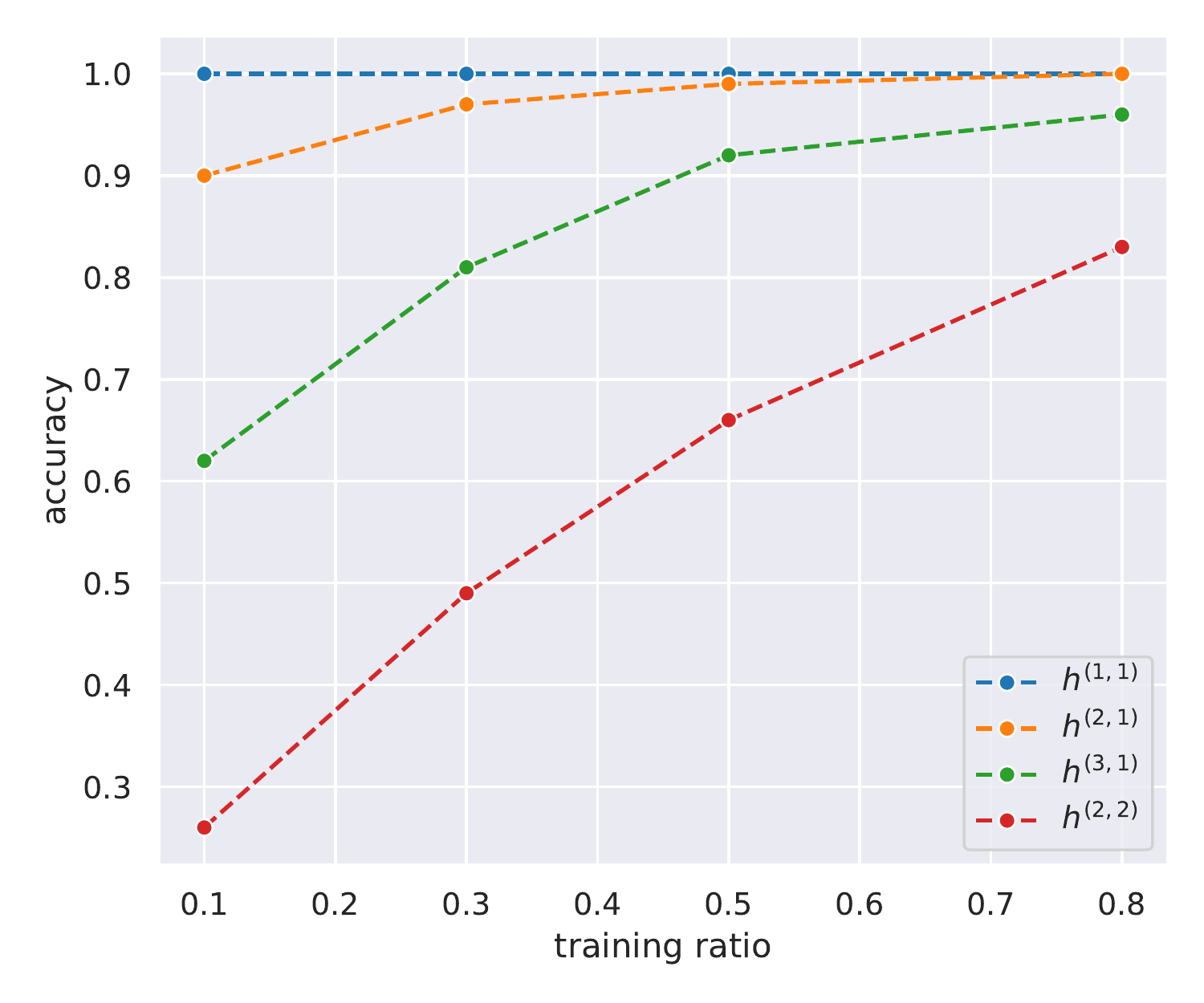}
    \hfill
    \includegraphics[width=0.47\textwidth]{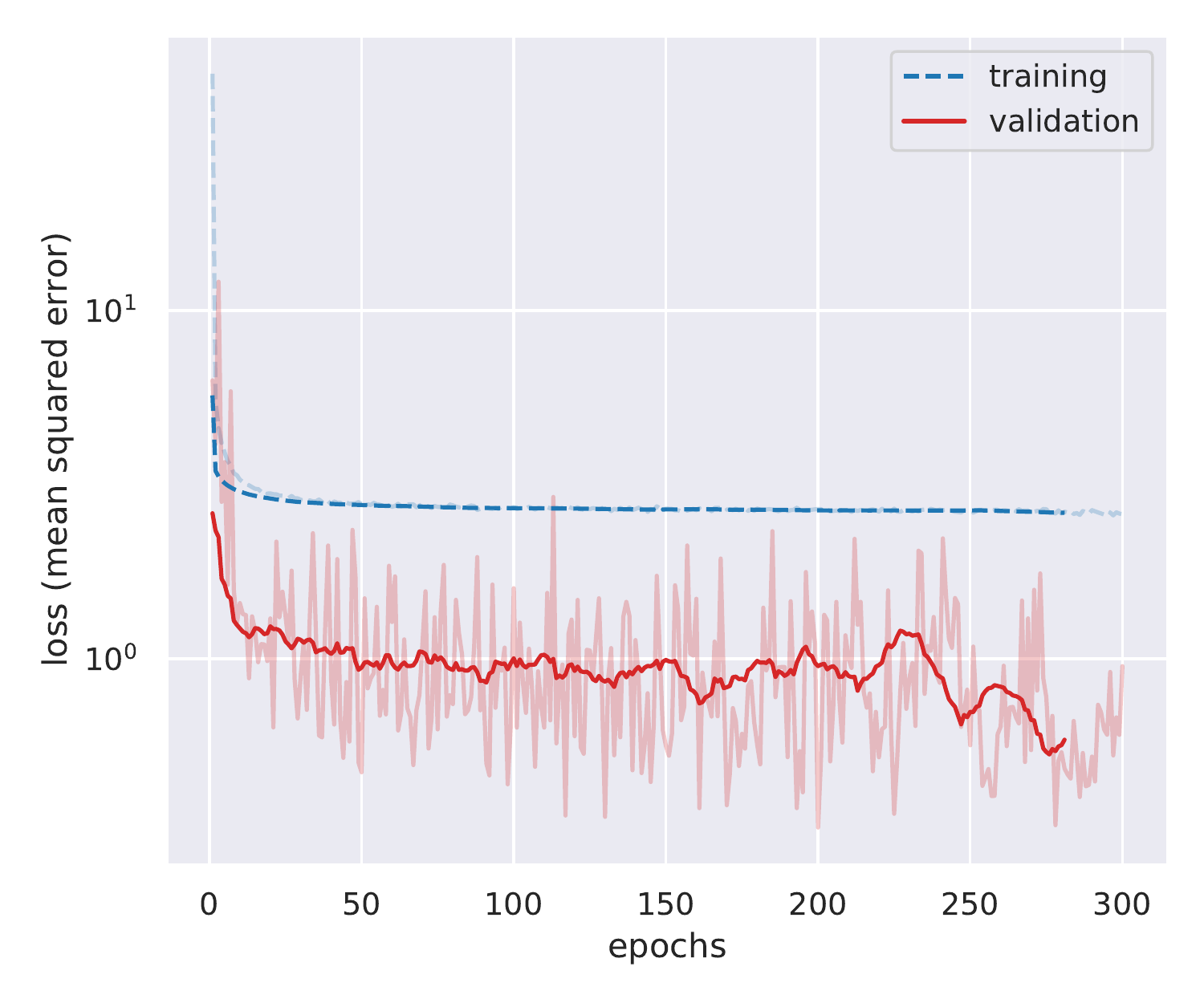}
    \end{minipage}
    \caption{\it On the left, we show the final test accuracy of CICYMiner for the four different training ratios. On the right, we present the loss function at \SI{80}{\percent} training ratio, smoothed with a running average over \num{20} epochs.}
    \label{fig:reg}
\end{figure*}

\subsection{Classifying Hodge numbers}
\label{sec:class}

Problems in image recognition are usually formulated as classification tasks. Take the {\it ImageNet} dataset which consists of $\num{14e6}$ data points with over $\num{21 000}$ classes. That is about one order of magnitude larger, both in samples and classes, than predicting $h^{(2,2)}$. In this section, we will train one neural network to classify each of the four non-trivial Hodge numbers independently. We will use an architecture based on Inception modules~\cite{Szegedy:2015:GoingDeeperConvolutions,Szegedy:2016:RethinkingInceptionArchitecture,Szegedy:2017:Inceptionv4InceptionResNetImpact} as was done for the best performing predictors of the CICY three-fold Hodge numbers~\cite{Erbin:2020srm,Erbin:2020tks}.
This specific architecture has been shown to lead to the best performance on the configuration matrices when using 1d kernels of maximal size. This partially reflects the fact that scanning coordinates in each projective space and a single variable over all projective spaces helps in better learning the connections between the different hypersurfaces of the CICYs (see~\Cref{fig:Inception}).
The choice of maximal 1d kernel is, in fact, motivated by the mathematical machinery required to compute Hodge numbers. There, one has to compute the dimension of ambient space cohomology group representations, which are stacked for each projective space. These ambient space representations arise after splitting up the Koszul resolution
\begin{align}
\label{seq:koszul}
    0  \to \wedge^K \mathcal{N}^* \to ... \to \mathcal{N}^* \to \mathcal{O}_\mathcal{A} \to \mathcal{O}_\mathcal{A}|_\mathcal{M} \to 0\; .
\end{align}
which contains the antisymmetric products $\wedge^s \mathcal{N}^*$ of the defining hypersurfaces ($\mathcal{N}$ denotes the normal bundle, which contains the information about the polynomial degrees $p^i_j$).
Moreover, using an Inception based architecture lead to a performance increase of misclassification rate on the {\it ImageNet} dataset from \SI{15.3}{\percent} for AlexNet~\cite{10.1145/3065386} using a standard convolutional architecture to \SI{6.7}{\percent} for the first version of GoogLeNet~\cite{Szegedy:2015:GoingDeeperConvolutions}.

We proceed as in earlier studies~\cite{Erbin:2020srm,Erbin:2020tks} by considering different train:val:test splits with respectively \SIlist{10; 30; 50; 80}{\percent} training and \SI{10}{\percent} validation data. The architecture hyperparameters have been optimized using Bayesian Optimization Hyperband~\cite{pmlr-v80-falkner18a,JMLR:v18:16-558} on the problem of predicting $h^{(3,1)}$. The same hyperparameters have then been used to also classify the other three Hodge numbers.

We opted to present the results of neural networks with a comparable number parameters $\num{840000} \pm \num{10000}$ to the number of configuration matrices. This architecture comprises four Inception modules, with respectively $3 \times 64$ and $16$ filters, utilizing batch normalization for better gradient propagation into the earlier layers~\cite{Szegedy:2015:GoingDeeperConvolutions,Szegedy:2016:RethinkingInceptionArchitecture,Szegedy:2017:Inceptionv4InceptionResNetImpact,DBLP:journals/corr/IoffeS15}. \Cref{fig:Inception} decomposes an Inception module into its different ingredients. The convolutional kernels scan over the configuration matrix dimensions, i.e.\ the maximal number of possible projective ambient spaces (16) and the maximal number of polynomial constraints (20). The Inception modules are followed by three dense layers with $16$ units, ReLU activation function and dropout layers with a \num{0.2} rate to contrast overfitting. Furthermore, we employ $\ell_1$ (\num{e-5}) and $\ell_2$ (\num{e-6}) regularization for all weights in the network. The last layer contains a softmax activation function with $\{h^{(i,j)}_{\text{min}}, \dots, h^{(i,j)}_{\text{max}} \}$ classes. The network is trained with Adam optimizer and an initial learning rate of \num{4e-4} on a $32$ mini-batch size.
This architecture is still trainable in a reasonable amount of time on a desktop computer with access to a GPU. In comparison to earlier studies~\cite{Erbin:2020srm,Erbin:2020tks}, we found that leaving the outliers inside the training data does not negatively impact the results.

\Cref{fig:class} shows in the top row the training loss and accuracy, and in the bottom row validation accuracy tracked over the training process and test accuracy for the best-performing model. The best-performing model is the one with the highest validation accuracy, which one would get when employing early stopping on that metric. It is important to track the best performing models as sometimes the loss starts increasing again as visible from the $h^{(2,1)}$ curve. The error bars are computed from the different training ratios and the budget on the $x$-axis is given by
\begin{multline}
    \text{budget}
        = \text{number of epochs} \\
            \times \frac{\text{percentage of training data}}{80}.
\end{multline}
We observe that $h^{(1,1)}$ is predicted with almost perfect accuracy for any training ratio, while the accuracies of the other three Hodge numbers improve with more training data. However, when the training data contains more than \SI{30}{\percent} of the samples one has diminishing returns for the accuracy. This is in line with previous observations for the CICY three-folds~\cite{Erbin:2020srm,Erbin:2020tks}. Even though the hyperparameters have been optimized to learn $h^{(3,1)}$, it is the worst performing value. This is interesting as Figure \ref{fig:Hodgehist} shows that the distribution of $h^{(2,2)}$ spans a longer range, contains more outliers and has a thicker tail. The plots show that we avoid overfitting to the training data.

\begin{table}[t]
    \centering
    \begin{tabular}{@{}ccccc@{}}
        \toprule
        & $h^{(1,1)}$ &
            $h^{(2,1)}$ &
            $h^{(3,1)}$ &
            $h^{(2,2)}$
        \\
        \midrule
        \SI{10}{\percent} &
            0.99 &
            0.87 &
            0.59 &
            0.62
        \\
        \SI{30}{\percent} &
            \textbf{1.00} &
            0.91 &
            0.67 &
            0.73
        \\
        \SI{50}{\percent} &
            \textbf{1.00} &
            0.94 &
            0.68 &
            \textbf{0.75}
        \\
        \SI{80}{\percent} &
            \textbf{1.00} &
            \textbf{0.95} &
            \textbf{0.70} &
            \textbf{0.75}
        \\ \midrule
        mean &
            1.00 &
            0.92 &
            0.66 &
            0.71
        \\
        \bottomrule
    \end{tabular}
    \caption{ \it Comparison of the test accuracy for different training ratios.}
    \label{tab:trainacc}
\end{table}

\Cref{tab:trainacc} collects the accuracy at different training ratios and the mean-value for the four different training ratios of the best performing model.\footnote{Using a five-fold increase in network weights (\num{4e6}) one is able to improve the accuracy of $h^{(3,1)}$ and $h^{(2,2)}$ to over \SI{80}{\percent}. However, this comes at the cost of significant more training time and we then enter the regime where there are more weights than samples in the dataset.}
In the training process, we employed learning rate decay with a factor of $0.4$, when the validation accuracy did not improve for epochs equivalent to \num{0.15} $\times$ budget. This is clearly visible from the loss and accuracy plots in the top row and accounts for the down- and up-stairs steps. Summarizing the results, we find that the hyperparameters found for predicting the worst performing Hodge number $h^{(3,1)}$ also generalize well to the other three Hodge numbers. This is a first indication that the prediction of Hodge numbers could benefit from multi-task learning.

\section{CICYMiner}
\label{sec:results}

In the previous section, we showed that a classification task based on Inception modules is effective in learning the Hodge numbers.
As the optimization was conducted for $h^{(3,1)}$, rather than an ad hoc structure for each output, the good results motivate further study on learning several Hodge numbers at the same time.
In this section, we focus on a regression model for two main reasons.
First, in general computations of vector bundle cohomologies, the predictions may not be bounded, thus an inference model has to be able to adapt by learning an approximation function, rather than classification probabilities.
Second, previous studies showed that regression models on a similar task were more efficient than classification~\cite{Erbin:2020srm}.

\Cref{fig:deepminer} shows the schematic of the architecture used in this section.
The architecture enables multi-task learning by hard parameter sharing over an initial structure capable of learning a shared representation of the input.
This, in general, has proven efficient at increasing the learning power of a single network, rather than differentiating and optimizing several, and to reduce the risk of overfitting~\cite{Caruana93multitasklearning:, baxter_bayesian/information_1997}.
The median layers of the network replicate a similar multi-tasked structure on the same learning objective: in fact, one branch of the sub-structures learning the Hodge numbers is an auxiliary architecture used to reinforce the stability of the representation.
No additional regularization was added to the model, apart from a \num{0.2} dropout rate before the fully connected networks in the auxiliary branches.
Such an architecture is thus capable of ``mining'' richer and more diverse features from a shared representation of the input by using different layer combinations.
The model is partly inspired by a recently proposed \emph{DeepMiner}~\cite{2021arXiv210209321B} model, used for people re-identification tasks, capable of learning more information by using different branched structures and layers.
As such, we refer to our model as \emph{CICYMiner}: we leverage the DeepMiner architecture with the advantages of multi-task learning in order to learn a family of related tasks, which however present complicated and strongly diverse distribution functions (see \Cref{fig:Hodgehist}).
The role of the auxiliary branches in CICYMiner (see~\Cref{fig:deepminer}) is mainly related to \emph{feature mining}, that is the ability to extract as much information as possible from intermediate representations, in order to guide the learning of the weights during learning. The auxiliary branches have, in fact, slightly different architectures with respect to the main branches, in order to perform different transformation on the inputs. An added value of the auxiliary branches is the duplication of the outputs, which in this multi-task context can improve overall performance, with regard to outlier and overfit control.

\subsection{Preprocessing and Evaluation Strategy}

We use the same dataset presented for the classification objective in the previous section.
Given the strong class imbalance, we select the training set by using a stratified approach on $h^{(2,1)}$ in order to preserve the distribution of the samples.
The validation set is then chosen totally at random, using \SI{10}{\percent} of the samples.
The remaining samples form the test set.
We preprocess the input data by simply rescaling the entries of the configuration matrices in the training set to the interval $[ 0,\, 1 ]$.
Matrices in the validation and test sets are rescaled accordingly, using the statistics obtained from the training set.

The outputs of CICYMiner are, in fact, floating point numbers $\tilde{h}^{(i,\,j)} \in \mathbb{R}^+$, as it is typical in regression tasks. They ultimately need to be rounded to integers to be directly compared with the true values and to compute the accuracy.
The distributions of the Hodge numbers have not been rescaled as training led to lower accuracy when this strategy was adopted.
The specialised branches of the network are, in fact, deep enough to apply the proper scaling starting from a shared representation and correctly learn the output distribution of the Hodge numbers.

In order to test the robustness and versatility of the network, we choose to keep the outliers in the training set.
In multi-task learning architectures, they may strongly affect the behavior of the network and may need robust loss functions during training~\cite{Zhang:2021:RobustMultiTask}: this problem is directly addressed in what follows.
On the other hand, what represents an outlier for a certain task, can be valuable information for another~\cite{Zhang:2021:SurveyMultiTask}, hence the choice of keeping the outliers in the training set.
Empirically, we also experienced a decrease in accuracy when trying to find a good outlier exclusion strategy.

\subsection{Training}

In this case, training occurred over a fixed amount of \num{300} epochs, due to time restrictions on the cluster computing infrastructure.
Training takes approximately \num{5} days on a single NVIDIA V100 GPU.
We use the Adam~\cite{kingma2014adam} stochastic gradient descent with an initial learning rate of \num{e-3} and a mini-batch size of \num{64} configuration matrices. 
Due to the long training time, the optimization was done using a grid search over a reasonable amount of choices of hyperparameters.
The network is ultimately made of \num{e7} trainable parameters, accounting for both the shared representation and the eight sub-networks learning Hodge numbers and their auxiliary outputs.
In terms of typical computer vision multi-task learning, we still deal with a small network: for instance, the original Inception network by Google has \num{0.7e7} parameters for a single classification task~\cite{Szegedy:2015:GoingDeeperConvolutions,Szegedy:2016:RethinkingInceptionArchitecture,Szegedy:2017:Inceptionv4InceptionResNetImpact}.

We already motivated the choice of keeping the outliers in the training set.
We address the arising issues by employing a \emph{Huber} loss function~\cite{10.1214/aoms/1177703732}:
\begin{equation}
    \small
    \mathcal{H}^{\{ k \}}_{\delta}( x )
    =
    \begin{cases}
        \displaystyle
        \frac{1}{2}
        \sum\limits_{n = 1}^k
        \sum\limits_{i = 1}^{N_k}\,
        \omega_n
        \big( x^{(i)} \big)^2,
        &
        \left| x^{(i)} \right| 
        \le \delta
        \\
        \displaystyle
        \delta
        \sum\limits_{n = 1}^k
        \sum\limits_{i = 1}^{N_k}\,
        \omega_n
        \left(
            \big| x^{(i)} \big| 
            -
            \frac{\delta}{2}
        \right),
        &
        \left| x^{(i)} \right| 
        > \delta
    \end{cases}
    \label{eq:huber}
\end{equation}
where $\omega_n$ for $n = 1, 2, \dots, k$ are the loss weights of the different branches of the CICYMiner, $\delta$ is a hyperparameter of the model and $x^{(i)}$ is the residual error of the $i$-th sample.
The choice of the loss turns out to be extremely useful in this regression task, as it behaves as a $\ell_2$ loss for small residuals, and it is linear for larger errors.
Robustness is thus implemented as a continuous interpolation between the quadratic and linear behaviour of the loss function.
This is a solution usually adopted for classification~\cite{Zhang:2021:SurveyMultiTask} where combinations of $\ell_1$, $\ell_2$ and Frobenius norm are used for robustness.

In our best implementation, we used $\delta = 1.5$, and loss weights \SIlist{0.05; 0.3; 0.25; 0.35}{} for $h^{(1,1)}$, $h^{(2,1)}$, $h^{(3,1)}$ and $h^{(2,2)}$, respectively (the auxiliary branches use the same values as the principal ones).
The learning rate is set to reduce by a factor of \num{0.3} after \num{75} epochs without improvements in the total loss of the validation set (as a reference, at \SI{80}{\percent} training ratio, this hard reduction mechanism triggered only once between epochs \num{270} and \num{300}).

\subsection{Results}

The final results are presented in \Cref{fig:reg} and in the last row of \Cref{tab:ablation}.
As shown in the learning curve, $h^{(1,1)}$ reaches perfect accuracy with just \SI{10}{\percent} of the training data, in alignment with previous attempts~\cite{He:2020lbz} and the classification results of the previous section.
$h^{(2,2)}$ is in general the most difficult label to train and it is strongly dependent on the training ratio.
The network appears to be underfitting the distributions of the Hodge numbers, and validation loss is still decaying after \num{300} epochs: it would be interesting to run training for longer time, in order to study the behaviour of the network.
At a training ratio of \SI{30}{\percent} the network reaches perfect accuracy on $h^{(1,1)}$, while $h^{(2,1)}$ gets to \SI{97}{\percent}. $h^{(3,1)}$ remains at \SI{81}{\percent}, while $h^{(2,2)}$ reaches barely \SI{49}{\percent}.
Increasing the number of training samples is, in general, beneficial for all Hodge numbers: $h^{(1,1)}$ and $h^{(2,1)}$ reach \SI{100}{\percent}, while the accuracy of $h^{(3,1)}$ and $h^{(2,2)}$ rises to \SI{96}{\percent} and \SI{83}{\percent}, respectively, when the training ratio reaches \SI{80}{\percent}.
For the first three outputs in~\Cref{tab:ablation}, the regression metrics, Mean Squared Error (MSE) and Mean Absolute Error (MAE), show the ability to effectively learn the discreteness of the Hodge numbers: both metrics show, in fact, values which can be confidently rounded to well defined integer results (i.e. $\text{MAE} \ll 0.50$ and $\text{MSE} \ll 0.25$).

The good performance of the first three Hodge numbers suggests the possibility to use relations such as the Euler characteristic \eqref{eq:euler}, which can be computed from combinatorics, and the linear constraint \eqref{eq:linear}.
Using the latter to compute $h^{(2,2)}$ leads to an accuracy of \SI{96}{\percent} on the test set, using the best results at \SI{80}{\percent} training ratio.
Using \eqref{eq:linear} and \eqref{eq:euler} together, $h^{(3,1)}$ and $h^{(2,2)}$ can reach perfect accuracy at \SI{80}{\percent} training ratio.
Using CICYMiner it is therefore possible to compute all four Hodge numbers with \SI{100}{\percent} of accuracy.

\begin{figure*}[t]
    \centering
    \begin{subfigure}[b]{0.47\textwidth}
        \includegraphics[width=\textwidth]{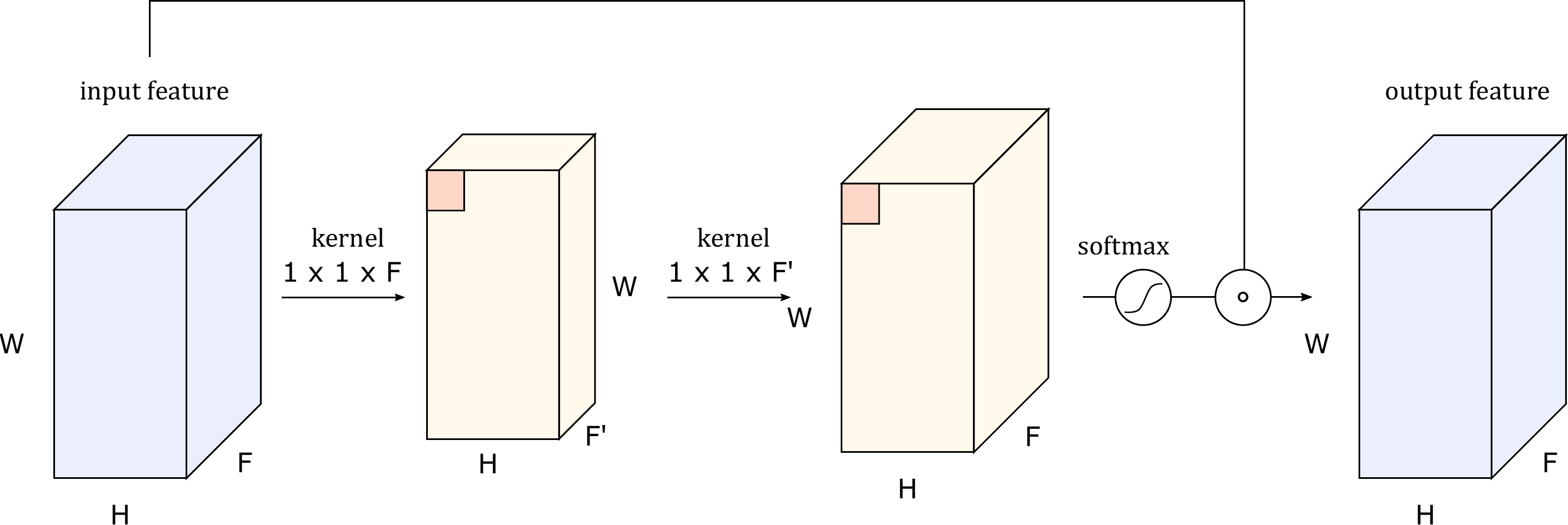}
        \caption{Channel Attention module.}
    \end{subfigure}
    \hfill
    \begin{subfigure}[b]{0.47\textwidth}
        \includegraphics[width=\textwidth]{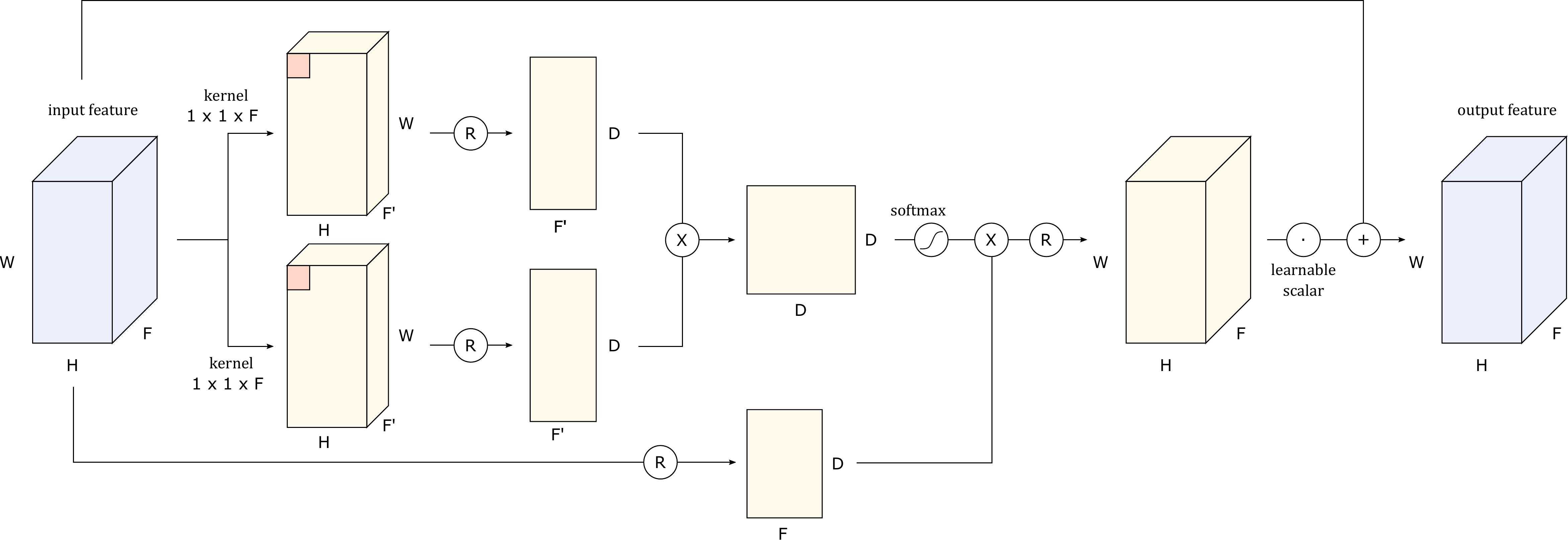}
        \caption{Spatial Attention module.}
    \end{subfigure}
    \caption{\it Substructures of the attention mechanism used in the ablation study. Here, $\times$ indicates a matrix product along appropriate axes, while $\circ$ is the Hadamard product (element-wise). Reshape operations (R) are also indicated.}
    \label{fig:attention}
\end{figure*}

\subsection{Ablation study}
\label{sec:ablation}

CICYMiner introduces new elements, with respect to previous attempts at predicting Hodge numbers of CICYs~\cite{Erbin:2020srm, He:2020lbz}, namely:
\begin{enumerate}
    \item Huber loss for robustness;
    \item auxiliary branches.
\end{enumerate}
In this section, we separately analyse each new aspect, together with other variations of the architecture.
Specifically, we analyse the impact of the batch normalization used in the Inception modules.
We also address the use of attention mechanisms~\cite{bahdanau2016neural}, used in the DeepMiner model, which in our case did not lead to an improvement in accuracy, but rather to a faster training.

\begin{figure*}[t]
    \centering
    
    \begin{minipage}[b]{0.47\textwidth}
        \centering
        \begin{tabular}{@{}lcccc@{}}
        \toprule
                   & $h^{(1,1)}$   & $h^{(2,1)}$   & $h^{(3,1)}$   & $h^{(2,2)}$   \\ \midrule
        +att       & \textbf{1.00} & 0.99          & \textbf{0.96} & 0.81          \\
        MSE loss   & \textbf{1.00} & 0.97          & 0.92          & 0.50          \\
        no aux     & \textbf{1.00} & 0.84          & 0.92          & 0.72          \\
        bs-256     & \textbf{1.00} & 0.99          & 0.94          & 0.65          \\
        layer norm & \textbf{1.00} & 0.99          & 0.92          & 0.66          \\ \midrule
        \textbf{CICYMiner}      & \textbf{1.00} & \textbf{1.00} & \textbf{0.96} & \textbf{0.83} \\
        \small{~~MSE (\num{e-4})} & \num{1.3}  & \num{98}  & \num{560}  & \num{6800}            \\
        \small{~~MAE (\num{e-3})} & \num{7.8}  & \num{19}  & \num{130}  & \num{360}             \\ \bottomrule
        \end{tabular}
        \captionof{table}{\it Comparison of the accuracy obtained by similar models at \SI{80}{\percent} training ratio. Regression metrics are also specified for CICYMiner at the same ratio.}
        \label{tab:ablation}
    \end{minipage}
    \hfill
    \begin{minipage}[b]{0.47\textwidth}
        \centering
        \includegraphics[width=\textwidth]{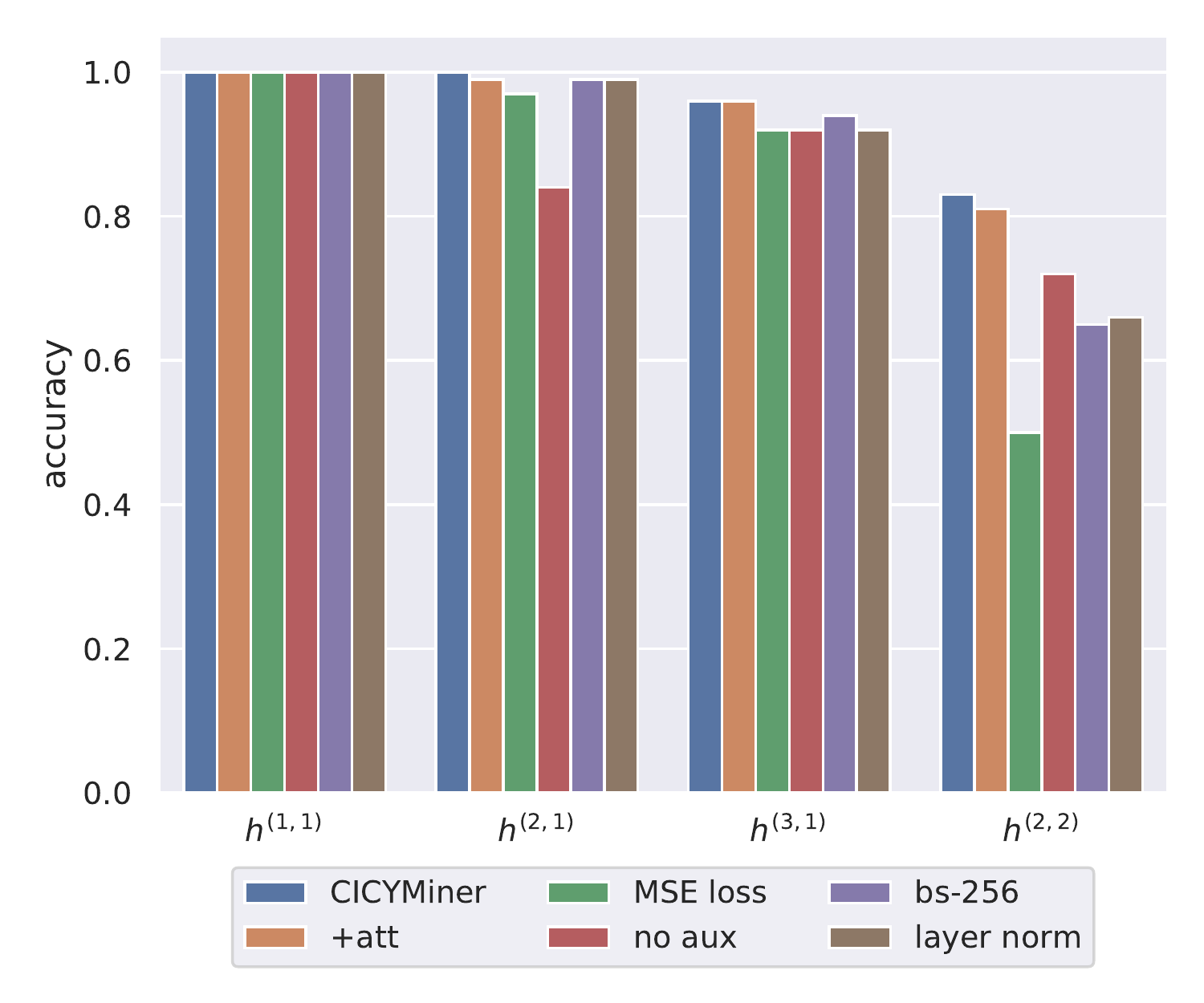}
        \caption{\it Summary of the ablation study.}
        \label{fig:ablation}
    \end{minipage}
\end{figure*}

\begin{figure*}[t]
    \centering
    \begin{subfigure}[b]{0.47\textwidth}
    \includegraphics[width=\textwidth]{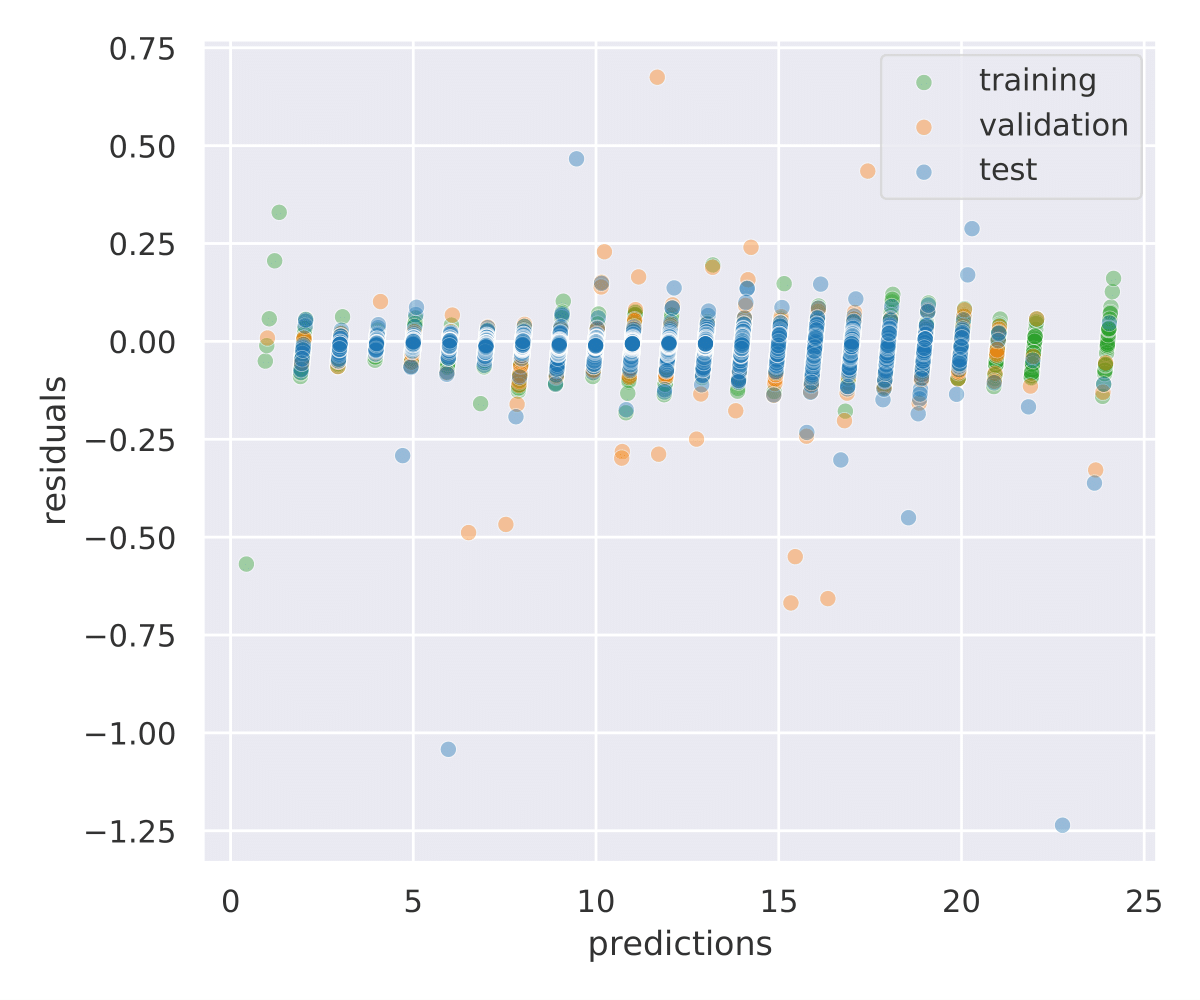}
    \caption{$h^{(1,1)}$ residuals.}
    \end{subfigure}
    \hfill
    \begin{subfigure}[b]{0.47\textwidth}
    \includegraphics[width=\textwidth]{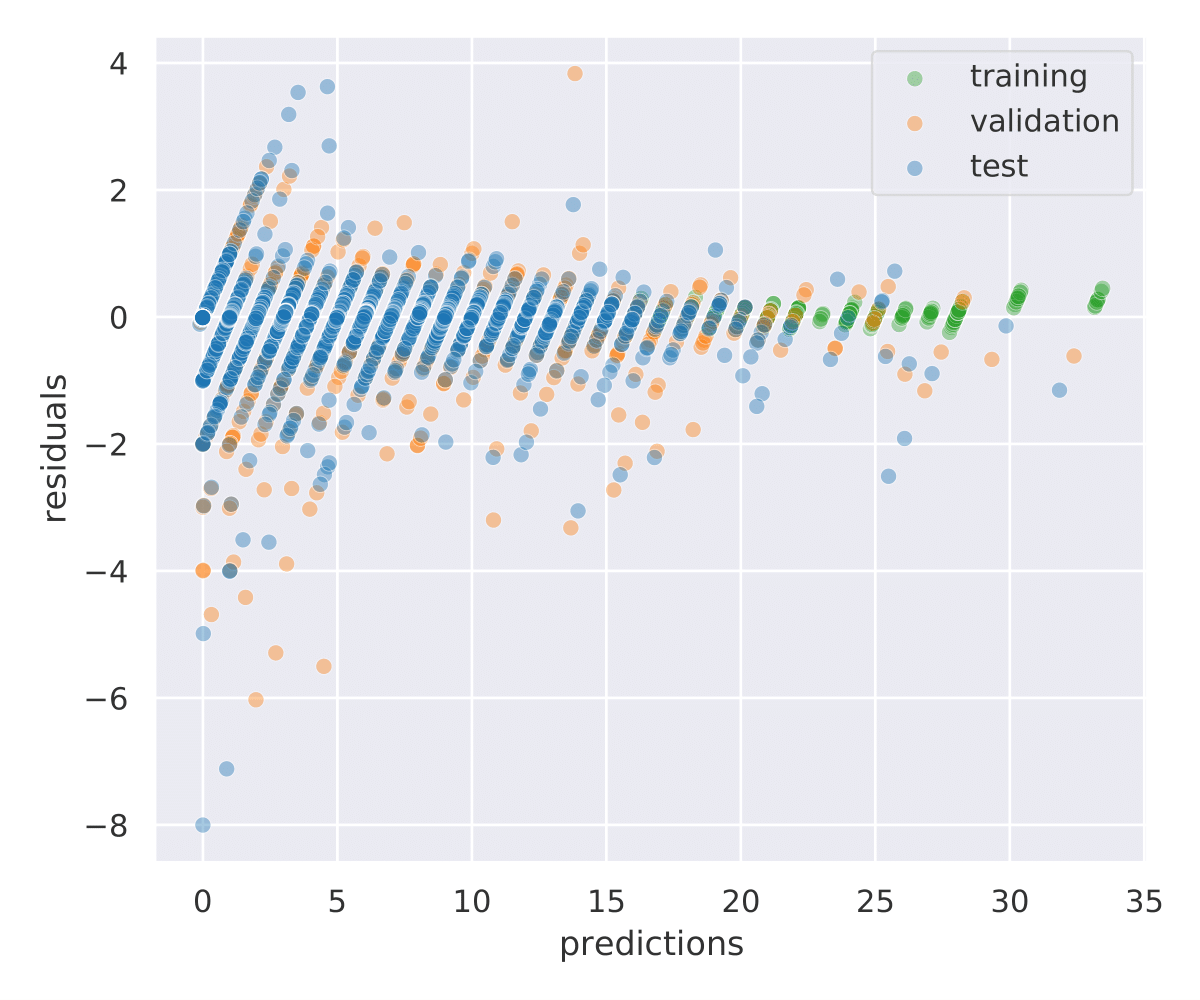}
    \caption{$h^{(2,1)}$ residuals.}
    \end{subfigure}
    \\
    \begin{subfigure}[b]{0.47\textwidth}
    \includegraphics[width=\textwidth]{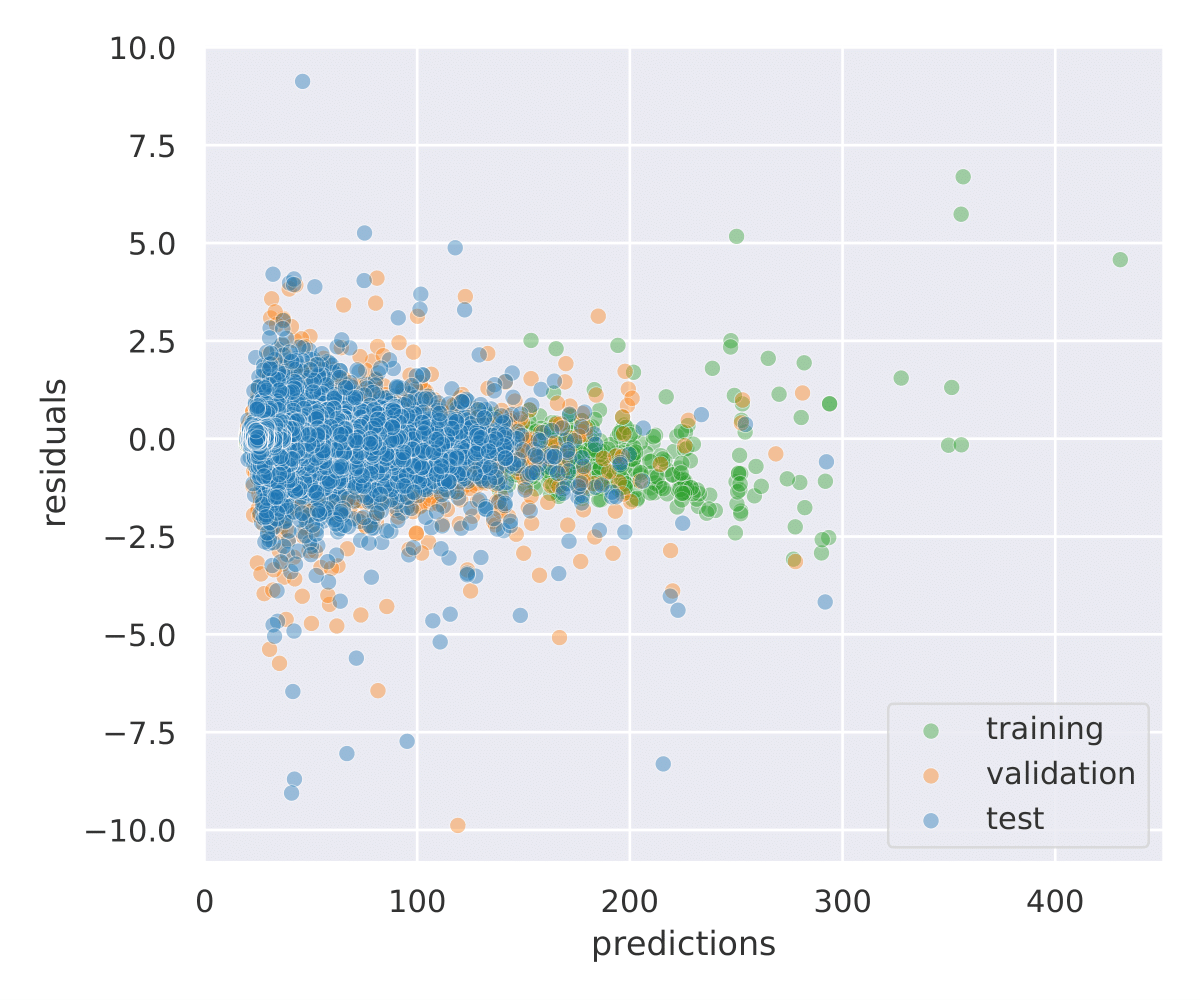}
    \caption{$h^{(3,1)}$ residuals.}
    \end{subfigure}
    \hfill
    \begin{subfigure}[b]{0.47\textwidth}
    \includegraphics[width=\textwidth]{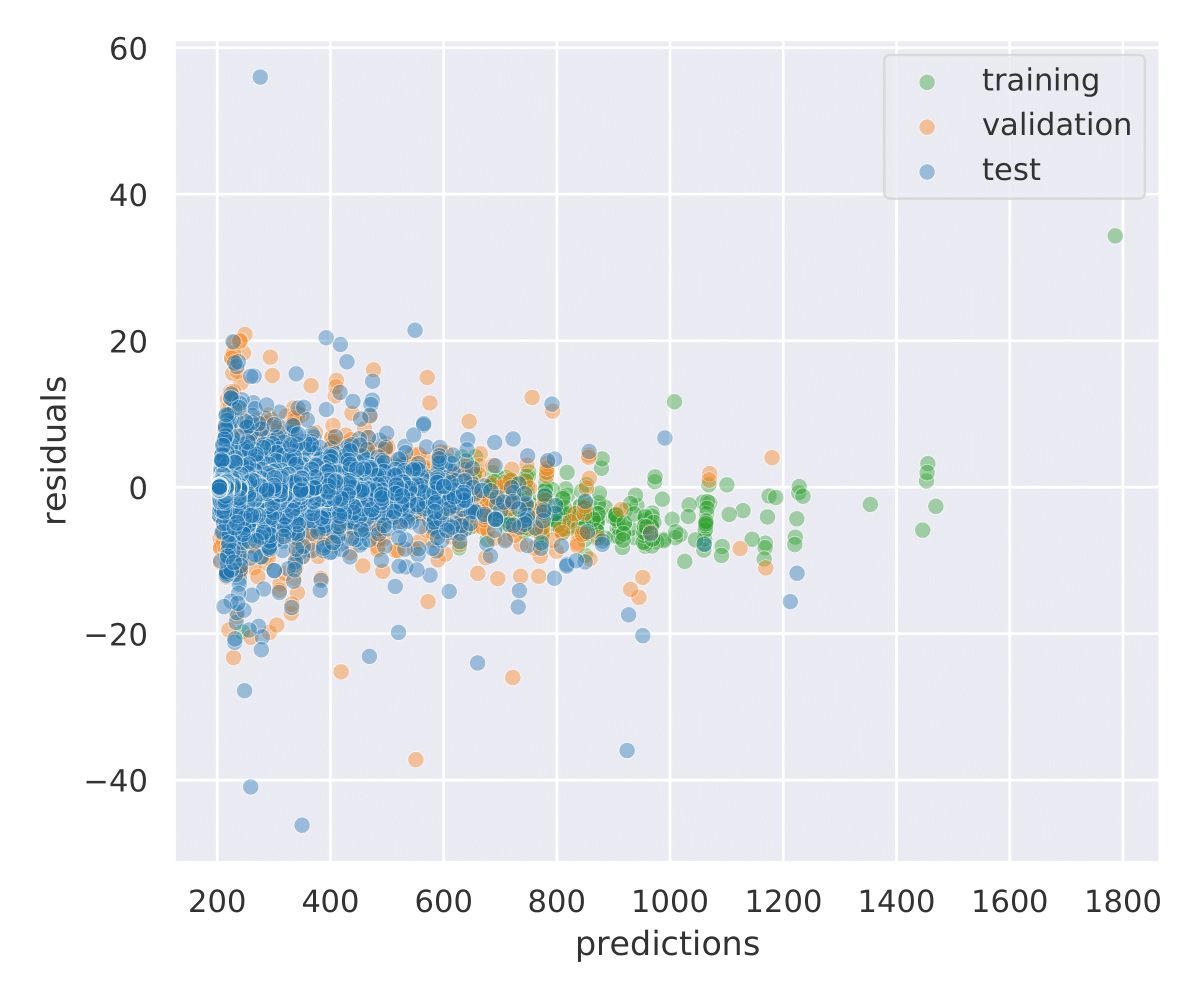}
    \caption{$h^{(2,2)}$ residuals.}
    \end{subfigure}
    \caption{\it Residual plots at \SI{80}{\percent} training ratio.}
    \label{fig:residual}
\end{figure*}

We proceed by modifying the backbone structure of CICYMiner.
We first introduce the attention mechanism used in~\cite{2021arXiv210209321B} for comparison.
The Spatial Attention Module ($\mathrm{SAM}$) and CHannel Attention Module ($\mathrm{CHAM}$) are presented in \Cref{fig:attention}: the full attention mechanism is the composition $\mathrm{CHAM} \circ \mathrm{SAM}$ used between each Inception module in the main branch of the task-specific architecture in \Cref{fig:deepminer}.
We also analyse the performance of the model by simply removing the auxiliary branches in the top layers of the network.
Then, as opposed to the Huber loss, we test the predictions using the usual MSE used in most regression tasks.
We finally change the size and type of the normalization strategy used in the architecture: we first train a network with a mini-batch size of \num{256} samples, and we then compare the results with a Layer Normalization~\cite{ba2016layer} strategy.
Results are summarised in \Cref{fig:ablation} and numerically reported in \Cref{tab:ablation}.
CICYMiner leads to the best overall performance for all four Hodge numbers. The distributions of the residuals, $x^{(i)}$ appearing in the Huber Loss~\eqref{eq:huber}, show in~\Cref{fig:residual} a homoscedastic behaviour (no correlations between predictions and absolute value of the residuals), which ultimately supports the completeness of the model and its ability to properly predict the four Hodge numbers correctly.

The use of a different loss function, which is not robust against outliers, led to largest drop in accuracy, overall: the difference starts to be consistent even for $h^{(2,1)}$ and $h^{(3,1)}$, which do not present many outliers in \Cref{fig:Hodgehist}.
The accuracy plummets when considering $h^{(2,2)}$, as expected.
The presence of outliers is also evident when increasing the mini-batch size: $h^{(2,2)}$ suffers the largest decrease in accuracy due to such normalisation strategy.
At the same time, the introduction of a batch-size independent Layer Normalization strategy, which normalizes each sample over the channel direction rather than the batch dimension, leads to a similar decrease.
The presence of outliers seems, therefore, a delicate issue for which the size of the mini-batches plays a relevant role.

A related aspect is represented by the ablation study on the auxiliary branches.
As their role is to mine a richer variety of features to stabilise the shared representation, and learn better approximations of the output, the accuracy drops significantly in the case of highly imbalanced distributions.
The largest drop impacts $h^{(2,1)}$ which suffers from predictions shifting towards zero.
This shows that we indeed need a mechanism to get as much training information as possible through the addition of transformations and auxiliary branches, as in the CICYMiner.

Finally, we analyse the impact of the attention modules: we insert such additional layers to improve the predictions of $h^{(3,1)}$ and $h^{(2,2)}$ only, as other Hodge numbers do not need additional transformations.
The results do not strongly differ from the case without the attention modules, though $h^{(2,2)}$ drops by \SI{2}{\percent} in accuracy.
It therefore seems that the attention modules do not help the predictions in this case, supported by the naive intuition that the configuration matrices do not suggest the development of a sequence model, such as in Natural Language Processing (NLP) or deep learning for video sequences.
However, the accuracy reached by the model occurs at around \num{100} training epochs, rather than \num{300} as in other cases.
The loss function then presents a slight increase after that.
The use of attention modules, together with an early stopping strategy, may therefore significantly cut the training time in this context.

\section{Conclusion}
\label{sec:outlook}

In this paper, we were able to show that Inception-based neural networks achieve good accuracy in predicting $h^{(3,1)}$ and $h^{(2,2)}$ and can reach perfect accuracy for the Hodge numbers $h^{(1,1)}, h^{(2,1)}$. Earlier studies using dense architectures were only able to work accurately with $h^{(1,1)}$~\cite{He:2020lbz}. Moreover, we showed that only a fraction of the training data is needed to already obtain promising results. This stands in contrast to earlier studies on CICY three-folds for which it was not possible to accurately predict $h^{(2,1)}$ (the only remaining non-trivial Hodge number in that case). The significant increase in dataset size is responsible for a good part of the increase in performance: the risk of overfitting is strongly reduced and generalization over all configuration matrices is more robust. This is also reflected in the observation that removing the tails of the Hodge number distribution is no longer needed in order to obtain good results. Our main results show that, given the two constraints \eqref{eq:euler} and \eqref{eq:linear} derived from tangent bundle indices, we are able to solve the problem of predicting all Hodge numbers with perfect accuracy.

Our results demonstrate that it is possible to obtain very accurate predictions for the dimension of cohomology groups with only partial training data. We emphasize that the computations of more generic vector bundle cohomologies also satisfy several linear relations and constraints derived from the index, Serre duality or vanishing theorems such as Kodairas. Thus, it is often sufficient to predict a single Hodge number with great accuracy to gain knowledge of all the others. In our experiments, training and validation error align, and we do not observe any significant high variance issues. The high validation and test accuracy suggests that the algorithm produces reliable results, even if it is only trained on partial data, say \SI{30}{\percent}. This should open up venues for further investigation into other vector bundle computations.

It is then important to find configurations which yield high accuracy on the validation set. In earlier studies, researchers have used feature enhancement to improve accuracy~\cite{Erbin:2020tks, He:2020lbz}. Unfortunately, it is not always possible to manipulate the input data via feature engineering in such a way. Adding a relevant monomial basis changes the dimension of the input space in non-trivial ways, such that one has to restrict oneself to a subset of the configuration matrices.

We opted to follow a model-centered approach, common in contemporary machine learning literature, by building a proper architecture with the right amount of parameters.
We balance the increased risk of overfitting, due to a larger number of trainable variables, with the natural regularization of multi-task architectures, thus increasing the final accuracy.
With its \num{e7} parameters, CICYMiner is, given its underlying geometric nature, still a small network with respect to many state-of-art models for computer vision or NLP. Good examples are represented by Inception-Resnet-v2~\cite{szegedy2016inceptionv4}, state-of-the-art in single-task image classification with \num{5.6e7} parameters and a long training time on \num{20} NVidia Kepler GPUs, and GPT-3~\cite{DBLP:journals/corr/abs-2005-14165}, state-of-the-art NLP model, with more than \num{175e9} model parameters.
In fact, recent research suggests that neural networks often admit power-scaling laws with dataset size and model parameters~\cite{DBLP:journals/corr/abs-2102-06701}.
It can also be noted that increasing the capacity of the model may be beneficial to the overall performance~\cite{belkin2019reconciling}. However, the geometric and physical interpretability might then become quite complicated and involved, hence the suggestion to constrain the complexity of the CICYMiner architecture.
It would be interesting to observe how far one can improve the accuracy of $h^{(3,1)}$ and $h^{(2,2)}$ by using larger networks or adding more data samples to the dataset, or even by just prolonging the training time on multiple GPUs.
Additional data samples can in principle be easily generated via (in-)effective splits of the already existing configuration matrices.
These redundant matrices had been discarded when compiling the initial dataset~\cite{Gray:2013mja}.

As a conclusion, our paper builds further the case for using deep learning in algebraic geometry by demonstrating that an appropriate neural network architecture can predict accurately Hodge numbers of CICY.
Moreover, since algebraic geometry uses datasets which are not of the type usually encountered in usual machine learning applications, our results extend their range of applications.

\section*{Acknowledgements}

RS is funded in part by the Swedish Research Council (VR) under grant numbers 2016-03873, 2016-03503, and 2020-03230. RS is grateful for financial support from the Liljewalch scholarship.
HE is funded by the European Union's Horizon 2020 research and innovation program under the Marie Skłodowska-Curie grant agreement No 891169.
HE is also supported by the National Science Foundation under Cooperative Agreement PHY-2019786 (The NSF AI Institute for Artificial Intelligence and Fundamental Interactions, \url{http://iaifi.org/}).
The work of RF is supported by a joint programme (PTC) between the \emph{Direction des énergies} and the \emph{Direction de la recherche tecnologique} of the CEA Paris--Saclay.
Computations were in part enabled by resources provided by the Swedish National Infrastructure for Computing (SNIC) at the HPC cluster \emph{Tetralith}, partially funded by the Swedish Research Council through grant agreement no.\ 2018-05973, and the \emph{FactoryIA} supercomputer, financially supported by the Ile-de-France Regional Council.

\printbibliography[heading=bibintoc]

\end{document}